\documentclass[twocolumn,showpacs,preprintnumbers,amsmath,amssymb]{revtex4}
\begin{document}
\title{Measuring Energy, Estimating Hamiltonians, and the Time-Energy
  Uncertainty Relation}
\author{Y. Aharonov}
\affiliation{School of Physics and Astronomy, Tel Aviv University,
Tel Aviv 69978, Israel.}
 \altaffiliation[Also at ] {Departement of Physics, University of
   South Carolina, Columbia,
SC 29208.}
\author{ S. Massar}
\affiliation{Service de Physique Th\'eorique,
Universit\'e Libre de Bruxelles, C.P. 225, Bvd. du Triomphe, B1050
Bruxelles, Belgium.}
\author{S. Popescu}
\affiliation{H.H. Wills Physics Laboratory, University of Bristol, Tyndall
Avenue, Bristol BS8-1TL, U.K.}
 \altaffiliation[Also at ] {
BRIMS, Hewlett-Packard Labs., Stoke Gifford,
Bristol  BS12-6QZ, U.K.}
\date{\today}
\begin{abstract}
Suppose that the Hamiltonian acting on a quantum system is unknown
and one wants to determine what is the Hamiltonian. We show that
in general this requires a time $\Delta t$ which obeys the
uncertainty relation $\Delta t \Delta H \gtrsim 1$ where $\Delta
H$ is a measure of how accurately the unknown Hamiltonian must be
estimated. We apply this result to the problem of measuring the
energy of an unknown quantum state. It has been previously shown
that if the Hamiltonian is known, then  the energy can in
principle be measured with arbitrarily large precision in an
arbitrarily short time. On the other hand we show that if the
Hamiltonian is not known then an energy measurement necessarily
 takes a minimum time $\Delta t$ which obeys the
uncertainty relation $\Delta t \Delta E \gtrsim 1$ where $\Delta E$
is the precision of the energy measurement. Several examples are
studied to address the question of whether it is
possible to saturate these uncertainty relations. Their interpretation
is discussed in detail.

\end{abstract}

\maketitle

\vspace{-1.1cm}

\section{Introduction}\label{Intro}

The uncertainty relations play a central role in quantum
mechanics. Their importance lies in the fact that they express in a
succinct manner the fundamental limitations on measurements imposed by
quantum mechanics. In particular it is these limitations that guarantee
that the mathematical formalism of the theory is free from
contradictions with experiment.

The time-energy uncertainty however has a particular status
because time is an external parameter in the theory and the energy
operator plays a special role since it determines the temporal
evolution. Thus the time energy uncertainty does not follow from the
commutation relations of two operators, but is determined indirectly,
for instance from the mathematical properties of the
Fourier transform with respect to the time variable.
The interpretation and status of the time-energy uncertainty
should therefore be examined with particular care.

In this paper we shall concentrate on energy measurements. By analogy
with other measurements, one expects that the time-energy uncertainty
expresses a fundamental constraint energy measurements must satisfy. The
aim of this paper is to clarify whether such a constraint exists, and
what should be its interpretation.

Suppose that one
must measure the energy of an unknown quantum state. One's first
intuition\cite{LP} in this case
is that
that the energy of an unknown state can be determined to accuracy
$\Delta E$ only if the duration $\Delta t$ of the measurement is
larger than
$1/\Delta E$ (since
$\Delta t \Delta E \geq 1$), 
where we set $\hbar = 1$.
Here and throughout,
by accuracy
of the measurement, we mean how much the result of the measurement
differs from the result of an ideal von Neumann measurement of the
Hamiltonian operator $H$. This will be defined with precision below.

Surprisingly, this intuition
is incorrect. It is possible to measure the energy of an unknown
quantum state
to arbitrarily high accuracy in an arbitrarily short time\cite{AB}. As an
illustration consider a spin 1/2 particle with a magnetic moment $\mu$
in a magnetic field $\vec B_0= B_0 \vec 1_z$ pointing along the $z$
direction. In order to measure the energy of the spin, one can
apply a strong magnetic field $\vec B(z) = B(z) \vec 1_z $
pointing in the $z$ direction with a gradient in the $z$
direction. This realizes a Stern-Gerlach  measurement of $\sigma_z$,
hence of the energy of the particle. The time necessary for this
measurement depends on the magnitude of the additional magnetic  field
$B(z)$, not on the original magnetic field $B_0$.
Since $ B(z)$ can be arbitrarily large, the
energy can be determined in an arbitrarily short time. Another
example, which is discussed in detail in the original article,
is the measurement of the energy
$H= p^2/2m$ of a free particle.

Thus the time-energy uncertainty seems not to apply
to energy measurements. Is this result universal? Or are there
cases where the time-energy uncertainty does apply and
measuring energy to an accuracy $\Delta E$ does
require a time $\Delta t$ limited by $\Delta t \Delta E \geq 1$?

We first note that if one attempts to devise an energy
measurement, it is often the case that this measurement will take
a time that satisfies the time-energy uncertainty. We refer
for instance to the example considered by Landau and
Peirels\cite{LP}, see also  Landau and Lifshitz\cite{LL}, in which
the kinetic energy of a particle is measured by letting it collide
with another particle initially at rest. In this example if the
energy is to be measured with precision, the initial momentum of
the particle at rest must be well known, but then its position is
spread out and the time at which the measurement takes place is
uncertain. Another well known example is the radiative decay of
an excited state. The emission of a photon by an excited atom that
decays to its ground state can be viewed as a measurement of the
energy of the electronic state of the atom. The mean time it takes
for the photon to be emitted, that is the lifetime of the excited
state, is then interpreted as the mean time it takes to make the
measurement. This lifetime is related to the line width, that is
to the uncertainty of the energy measurement, by the time-energy
uncertainty. Following \cite{AB} one would argue that such energy
measurements, which satisfy the time-energy uncertainty, are
simply badly designed, and that one can in principle make energy
measurements that do not obey the time-energy uncertainty.
Nevertheless it is surprising that  so many energy measurements do
obey the time-energy uncertainty. We would like to know what makes
some energy measurements less efficient than others.

There are also some particular
circumstances in which the time-energy uncertainty
must be obeyed in an energy  measurement.
Specifically consider the situation
in which one wants to measure
the energy of an isolated system. Since it is isolated,
ie. uncoupled to any exterior degrees of freedom, the measuring
apparatus must be internal to the system itself. In \cite{AR} it is
argued that this constraint
implies that measuring the total energy of an
isolated system
to accuracy $\Delta E$ requires a time $\Delta t$ that satisfies
$\Delta t \Delta E \geq 1$. Contrary to the examples mentioned in
the preceeding paragraph, where a more sophisticated strategy could in
principle measure the energy in a smaller time than that given by the
uncertainty relation, in this case the uncertainty relation must be
obeyed.

Thus the present status of the time-energy uncertainty in the
context of energy measurements is unsatisfactory. Different
examples suggest conflicting interpretations.

As we shall prove bellow, the resolution of this puzzle is the
following. When the Hamiltonian of the system is {\it known} the
conclusions of \cite{AB} hold, namely we can measure the energy as
precisely as we want in a time as short as we want. On the other
hand, {\it whenever the Hamiltonian of a system is completely
unknown,
determining what is the Hamiltonian to precision $\Delta H$
requires a time $\Delta t$ given by $\Delta t \Delta H \geq
1 $.}

The origin of the conflicting interpretations of the time-energy
uncertainty relation in the context of estimating the energy of an
unknown state is now clear. In some cases the energy measurement
can also serve to estimate an unknown Hamiltonian. In these cases,
because the measurement can serve this dual purpose, it requires a
time $\Delta t$ which is limited by the accuracy with which it
could estimate an unknown Hamiltonian. On the other hand the
measurements envisaged in \cite{AB} cannot estimate an unknown
Hamiltonian. These measurements measure an operator $A$
($\sigma_z$ in the example above, or the momentum $p$ in the
example of \cite{AB}).
 If the Hamiltonian is a function of
this operator $H=f(A)$ (for instance
$H= \mu B_0 \sigma_z$ or $H= p^2/2m$),
then the measurement of $A$ can be
used to determine the energy of an unknown state.
Such a von Neumann measurement of an operator $A$  can be realized in
an arbitrarily short time. But if the time is very short, the
measurement is brutal, that is the interaction between the measuring
apparatus
and the system dominates the evolution. It is therefore impossible to use
such a measurement to estimate an unknown Hamiltonian since the
action of the original Hamiltonian is completely masked by the
interaction.

Therefore in order to understand the interpretation of the time-energy
uncertainty relation in the context of energy measurements, it is
mandatory to first understand the limitations quantum mechanics
imposes on the estimation of the Hamiltonian acting on a system when
the Hamiltonian is unknown.
This problem is in a certain sense the
dual of the standard problem of quantum information in which one
must estimate the state of a quantum system. Here one must
estimate the dynamics. This is a fundamental problem in quantum
mechanics. In fact many experimental situations can be phrased in
this language. For instance there could be an unknown force acting
on a particle which one wants to estimate. Or the Hamiltonian
could depend on an unknown parameter which one wants to estimate.

We show in the present paper
that the precision with which one can estimate the
dynamics and the time required for this estimation are related by
an uncertainty relation
\begin{equation}
\Delta t \Delta H \geq 1 \label{DtDH}
\end{equation}
where the precise meaning of $\Delta H$
depends on the details of the problem.

It is intuitively obvious that the dynamics of a quantum
system cannot be estimated
  instantaneously. Indeed suppose that initially the state is
  $\psi_0$. Then after a time $t$ the state has evolved to
$\psi(t,H)=e^{-itH}\psi_0 \simeq \psi_0 + i t H \psi_0$. From this
expression it is clear that in order to estimate the dynamics, ie.
to estimate $H$, sufficient time must elapse that the zero'th
order term $\psi_0$ in the expansion in $t$ does not dominate.
Where this the case, the states $\psi(t,H)$ would be almost
identical to $\psi_0$ and hence undistinguishable. The results
reported in this paper make this qualitative statement precise.

The problem of estimating the dynamics has been considered
recently by Childs, Preskill and Renes\cite{CPR}. Their results
constitute a starting point for our discussion.

The remainder of the article is organized as follows. We first
discuss the problem of distinguishing with certainty
between two
Hamiltonians. Next we consider the case where one only wants to
estimate with finite error probability which of the two Hamiltonians
is the true Hamiltonian. Then we turn to the problem of estimating
what is the Hamiltonian when one has no prior knowledge about the Hamiltonian.
Finally we go
back to the problem of estimating
the energy of an unknown state. We discuss how it is related to
estimating an unknown Hamiltonian. We prove that if the Hamiltonian is
completely unknown then the time necessary to estimate the energy and
the precision with which the energy is estimated must
obey a time-energy uncertainty relation. We illustrate the problem of
estimating the energy of a state when the Hamiltonian is completely or
partially unknown by several examples.

\section{Distinguishing between two Hamiltonians}\label{EstH}

\subsection{Minimum time necessary to distinguish between two unknown
  Hamiltonians}\label{TwoH}

In this section we consider the problem of estimating the Hamiltonian
acting on a system when the Hamiltonian is unknown. We
consider in this section the special case in which there are
only two possible Hamiltonians that could act on the system, $H_1$ or
$H_2$, and one must distinguish with certainty which it is.

This particular problem
has been considered previously by Childs, Preskill and Renes\cite{CPR}.
They show that the
minimum time $\Delta t$ to determine with certainty
the Hamiltonian
must satisfy the constraint
\begin{equation}
\Delta t\  D_0(H_1,H_2) \geq  \pi  \ .
\label{TDH}
\end{equation}
where $D_0(H_1,H_2)$ measures how much the two Hamiltonians differ and is
defined as follows. Take the difference of the two Hamiltonians
$H^d= H_1 - H_2$. Denote by
$E^d_{max}$ and $E^d_{min}$ the largest and smallest eigenvalue
of $H^d$ respectively. Then $D_0(H_1,H_2)$ is given by
\begin{equation}
 D_0(H_1,H_2)
= \max \{ E^d_{max} - E^d_{min} ,|E^d_{max}|,  |E^d_{min}| \} \ .
\label{DH}
\end{equation}
(This expression generalises a result of \cite{CPR}).

In the appendix it is shown that
$D_0$ is a distance on the space of Hamiltonian
operators (ie. it is symmetric, positive and equal to zero only if the
two Hamiltonians coincide, and it obeys the triangle inequality). This
important property is central to some of the arguments below.

In order to make the problem considered by Childs,
Preskill and Renes
more concrete, consider the following example.
You are given a box in which there is one of two unknown
magnetic fields $\vec B_1$ or $\vec B_2$. Your task is to determine
which kind of box you have. The only way you can probe the box is to
send through the
box a specific kind of particle of magnetic moment $\mu$. Thus the two
boxes differ in that they act as two different Hamiltonians $H_{1,2}
 = \mu
\vec B_{1, 2} \cdot \vec \sigma$.
Given sufficient time or a sufficient supply of
particles you can always determine what is the magnetic field.
(We suppose that the
time that the particles pass in the box can be freely chosen, for instance
by choosing their initial velocity. And one can send the particle
through the box as many times as one wants). However if you are
given only one particle, what is the minimum time necessary to
accomplish this task? Childs, Preskill and Renes show that this time
is limited by eq. (\ref{TDH}).

We now present a proof of eqs. (\ref{TDH}, \ref{DH}) that
is inspired by the techniques developed to study
interaction free measurements in \cite{MM}.
The proof is also closely connected to the bounds on oracle query
  complexity obtained in \cite{BBBV} and even more so to a continuous-time
  analogue obtained in \cite{FG}.
We first describe the most general strategy that could be used.
First of all, we consider that we may let the particle go into
the box or not, i.e. for certain values of its position the
particle can pass through the box and for others not. This allows
us to make a superposition of particle passing through the box and
not passing through the box. Mathematically we describe this by
the decomposition of the Hilbert space into the sum ${\cal H} =
{\cal H}_{box} \oplus {\cal H}_{Nbox}$ where ${\cal H}_{box}$
corresponds to the particle passing through the box and ${\cal
H}_{Nbox}$ corresponds to the particle passing next to the box.
Second, we consider that the particle can also be entangled with
some other particle, called ancilla, onto which the hamiltonian
doesn't act (eg. the ancilla is kept out of the box). The Hilbert
space is thus further increased to ${\cal H} = ({\cal H}_{box}
\oplus {\cal H}_{Nbox})\otimes{\cal H}_{ancilla}$.

In the particular problem considered in this section, namely,
distinguishing between two possible Hamiltonians, using an ancilla
turns out to be irrelevant. (However when one must distinguish between
more  than two Hamiltonians, using an ancilla can be helpfull, see
\cite{CPR} for a simple example).
To simplify the proof we will first
consider the case with no ancilla and then show in section
\ref{ancilla}
that the presence
of the ancilla makes no difference.

 The most general strategy consists of sending the particle
several times through the box and making an arbitrary unitary
transformation on the particle after each passage. We describe
this as follows. Initially the particle is in state $|\psi_0
\rangle$. Before the k-th passage through the box the particle is
in state $|\psi_{k-1}^i\rangle$ where the superscript $i=1,2$
corresponds to which Hamiltonian $H_{1,2}$ is acting on the
particle. We decompose the state as $|\psi_{k-1}^i\rangle =
|u^i_{k} \rangle + | v^i_{k} \rangle$ where $| u^i_{k} \rangle \in
{\cal  H}_{box}$ and $| v^i_{k} \rangle \in {\cal  H}_{Nbox}$.
After passing through the box the particles state is
$|\psi_{k}^{'i}\rangle = e^{-i H_i t_k} | u^i_{k} \rangle
+|v^i_{k} \rangle$ where $t_k$ is the time the particle is in the
box (and we have supposed without loss of generality that if the
particle does not pass through the box the Hamiltonian is zero).
We then make an arbitrary unitary transformation on the particle
$|\psi_{k}^{'i}\rangle \to |\psi_{k}^i\rangle =
U_k|\psi_{k}^{'i}\rangle $. We can describe the whole evolution
succinctly by
\begin{equation}
|\psi_{k}^i\rangle = U_k  e^{-i \tilde H_i t_k} |\psi_{k-1}^i\rangle
\end{equation}
where $\tilde H_i$ is the extension of $H_i$ to
${\cal H}_{box} \oplus {\cal
  H}_{Nbox}$. It is equal to $H_i$ on $ {\cal H}_{box}$ and equal to
zero on ${\cal
  H}_{Nbox}$.

We therefore have
\begin{equation}
\langle \psi_{k}^1 | \psi_{k}^2 \rangle =
\langle \psi_{k-1}^1 | e^{i \tilde H_1 t_k}e^{-i \tilde H_2 t_k} | \psi_{k-1}^2
\rangle
\ ,
\end{equation}
or expressed differently
\begin{equation}
\langle \psi_{k}^1 | \psi_{k}^2 \rangle
- \langle \psi_{k-1}^1 | \psi_{k-1}^2 \rangle
=
\langle \psi_{k-1}^1 | e^{i \tilde H_1 t_k}e^{-i \tilde
H_2 t_k} -1 | \psi_{k-1}^2
\rangle
 \ .
\end{equation}
The procedure to distinguish between $H_1$ and $H_2$ can only improve if
the
total time $T$ is kept fixed, but the particle is allowed to pass more
times through the box. That is the discrimination can only improve if
one takes smaller time intervals $t_k$. In the limit of infinitesimal
$t_k$ one obtains
\begin{equation}
{ d \langle \psi^1 | \psi^2 \rangle \over dt}  = i
\langle \psi^1 |  (\tilde H_1 - \tilde H_2)|\psi^2 \rangle =  i
\langle \psi^1 |  \tilde H^d|\psi^2 \rangle
\end{equation}
where $\tilde H^d= \tilde H_1 - \tilde H_2$ is equal to $H^d= H_1-H_2$
on $ {\cal H}_{box}$ and equal to
zero on ${\cal
  H}_{Nbox}$.
The absolute value of the overlap therefore changes as
\begin{equation}
{ d | \langle \psi^1 | \psi^2 \rangle|^2 \over dt}
 = 4
Im [
\langle \psi^2 |\psi^1 \rangle \langle \psi^1 |  \tilde H^d|\psi^2 \rangle ]
\ .
\label{7}
\end{equation}
We can always write the states $| \psi^{1,2} \rangle$ as
\begin{eqnarray}
&|  \psi^{1} \rangle  =  e^{ +i \chi /2} \left (\cos (\theta/2)
 |\psi^\parallel\rangle + \sin (\theta /2) |\psi^\perp\rangle
\right )& \ ,\nonumber\\
&|  \psi^{2} \rangle  =  e^{ -i \chi /2} \left (\cos (\theta/2)
 |\psi^\parallel\rangle - \sin (\theta /2) |\psi^\perp\rangle
\right ) & \ ,\nonumber\\
&0 \leq \theta \leq \pi/2 \quad , \quad \langle\psi^\parallel
|\psi^\perp\rangle =0 &
\ .\end{eqnarray}
This enables us to write
(\ref{7}) as
\begin{equation}
{ d \cos^2 \theta \over dt}
 = 2 \cos \theta \sin \theta Im [ \langle \psi^\parallel | \tilde H^d |
\psi^\perp \rangle ]
\end{equation}
or equivalently
\begin{equation}
{ d \theta \over dt}
 =  Im [ \langle \psi^\parallel | \tilde H^d |
\psi^\perp \rangle ] \ .
\label{ddd}
\end{equation}
Now we note that for any operator $A$,
\begin{equation}
| Im [ \langle \psi^\parallel | A |
\psi^\perp \rangle ]|\leq {a_{max} -a_{min} \over 2}
\end{equation}
where $a_{max}$ is the largest eigenvalue of $A$ and $a_{min}$ the
smallest. Equality is attained if
\begin{eqnarray}
| \psi^\parallel \rangle &=&
{e^{i\varphi}|a_{max} \rangle + e^{i\varphi'}| a_{min} \rangle \over
  \sqrt{2}} \nonumber\\
| \psi^\perp \rangle &=& {e^{i(\varphi + \pi/2)} |a_{max} \rangle  +
e^{i(\varphi'-\pi/2)}
| a_{min} \rangle \over
  \sqrt{2}}
\label{psipsi}
\end{eqnarray}
where $\varphi$ and $\varphi'$ are arbitrary phases.
In the case of $\tilde H^d$, its largest eigenvalue is $\max\{E^d_{max},0\}$
and its smallest eigenvalue is $\min\{E^d_{min},0\}$
(where $E^d_{max}$ and $E^d_{min}$ are the largest and smallest
eigenvalues of $H^d$). Hence the
absolute value of the right hand side of (\ref{ddd}) is bounded by
$| Im [ \langle \psi^\parallel | \tilde H^d |
\psi^\perp \rangle ] | \leq D_0(H_1,H_2) /2$ where $D_0$ is given by
(\ref{DH}).
We therefore have
\begin{equation}
|{ d \theta \over dt}| \leq    D_0(H_1,H_2) / 2 \ .
\label{14}
\end{equation}
Integrating we have
\begin{equation}
-{  D_0(H_1,H_2) t\over 2} \leq \theta(t) - \theta(0)
\leq {  D_0(H_1,H_2) t \over 2}  \ .
\end{equation}
Initially $\theta(0)=0$ since $|\psi(0)\rangle$ is
independent of $i$. Requiring that one can recognize with certainty
which is the Hamiltonian, that is requiring $\theta(t) = \pi/2$,
one finds that
$t  D_0(H_1,H_2) \geq \pi $, as announced.

In the above proof the possibility of letting the particle go
through the box as well as outside the box allows us to extend the
Hamiltonian $H^d$ so that it also has the eigenvalue zero. This is
described by replacing $H_{i}$ with $\tilde H_{i}$. It is the
possibility that allows the maximum in eq. (\ref{DH}) to be taken
not only over the first term $E^d_{max} - E^d_{min}$, but also
over the two other terms, $|E^d_{max}|$ and $|E^d_{min}|$ and
makes our result eq. (\ref{DH}) differ from the result obtained in
\cite{CPR}. Indeed if one does not allow for this then eq.
(\ref{TDH}) continues to hold, but eq. (\ref{DH}) is replaced by
$D_0(H_1,H_2) =  E^d_{max} - E^d_{min}$. Thus in this case the
measurement may be less efficient.

In order to illustrate how the ``in/out of the box"
possibility can be used, we consider the following example. We are
given one of two black boxes, and must distinguish which box we
have. The boxes are conducting and hollow. They are connected to
an external electrostatic potential. The only way the two boxes
differ is that the potential can take two different values
$\phi_1$ and $\phi_2$. The only way we can probe which box we have
is by sending a charged particle through the box. The particle
will therefore acquire a phase which depends on the potential.
Mathematically we can describe this as the problem of
distinguishing two Hamiltonians $H_{1,2}=H_0 + \phi_{1,2} I$ where
$I$ is the identity operator. In order to distinguish which box we
have, we put the particle in a superposition of two states, one of
which passes through the box and the other which does not. Thus we
prepare the particle in the state $(|\mbox{
  through box}\rangle + |\mbox{Not through box}\rangle)/\sqrt{2}$ and
send the particle at a speed such that it passes a time
$T=\pi/ (\phi_1 - \phi_2)$ in the box. After this time the particle is
in one of two orthogonal states $(|\mbox{ Not
  through box}\rangle \pm e^{i\pi \phi_1/(\phi_1 - \phi2)}
|\mbox{through box}\rangle)/\sqrt{2}$
which can easily be distinguished.
On the other hand, if we had not been allowed to use the ``out of the
box'' alternative, we could not have distinguished between the 2
Hamiltonians since they differ only by a constant term which adds an
unobservable phase to the wave function.

There are also situations where it is not necessary to use the
``in/out of the box" possibility. Suppose one must distinguish,
using a spin 1/2 particle of magnetic moment $\mu$, between two
magnetic fields of equal magnitude but pointing in opposite
directions $B_{1,2} = \pm B_0 \vec 1_z$. The strategy in this case
is simply to prepare the spin in state $|\uparrow_x\rangle$ and
let it evolve in the magnetic field a time $T= \pi /\mu B_0$ and
then to measure the operator $\sigma_y$.

\subsection{Distinguishing between two time dependent Hamiltonians}
\label{timed}

In the preceeding section we considered the case where the two
Hamiltonians
which
must be distinguished are time independent. One can easily generalize
this result to the case where the Hamiltonians are time
dependent.

Let us suppose one must distinguish between two time dependent
Hamiltonians $H_1(t)$ and $H_2(t)$. The arguments of section
\ref{TwoH} can be followed unchanged until eq. (\ref{14}) which
becomes
\begin{equation}
|{ d \theta \over dt}| \leq   D_0(H_1(t),H_2(t)) / 2
\label{14B}
\end{equation}
where $D_0(H_1(t),H_2(t))$ is the instantaneous value of the distance
eq. (\ref{DH}). Integrating we have
\begin{eqnarray}
- \int_0^t dt \ {D_0(H_1(t),H_2(t)) \over 2} &\leq& \theta(t) - \theta(0)
\nonumber\\
&\leq&  \int_0^t dt \ {D_0(H_1(t),H_2(t)) \over 2} \ . \nonumber\\
\end{eqnarray}
Initially $\theta(0)=0$ since $|\psi(0)\rangle$ is
independent of $i$. Requiring that one can recognize with certainty
which is the Hamiltonian, that is requiring $\theta(t) = \pi/2$,
one finds that
\begin{equation}
\label{general}
\int_0^t dt \ D_0(H_1(t),H_2(t)) \geq \pi
\end{equation}
which is the generalization of the constraint eq. (\ref{TDH}) to the case
of time dependent Hamiltonians.

\subsection{Using an ancilla does not help to distinguish between two
Hamiltonians}\label{ancilla}

In the proof of eqs. (\ref{TDH}, \ref{DH})
and (\ref{general}) given in sections \ref{TwoH} and \ref{timed} we
did not consider the possibility that the 
particle passing through the box is entangled with another
particle (ancilla) was not considered. We shall now show that 
if we assume the ancilla is not allowed into the box,
then eqs. (\ref{TDH}, \ref{DH}) and (\ref{general}) continue to hold. 

To this end it is helpfull to view the particle and ancilla as a
single larger system with Hamiltonian $H_i^{total}= 
H_i(t)\otimes H_{ancilla}(t)$
where $H_i$ is unknown and $H_{ancilla}$ is known but arbitrary, and
to reformulate the task as the problem of distinguishing $H_1^{total}$
from $H_2^{total}$. Note that this reformulation englobes apparently
more involved strategies where for instance the ancilla is repeatedly
measured and the evolution made conditional on the results of these
intermediate measurements. Indeed by including the measuring device in
an even bigger ancilla one recovers the above formulation.

Now the distance $D_0$ between two such Hamiltonians obeys
\begin{eqnarray}
&D_0(H_1^{total}(t) , H_2^{total}(t))&\nonumber\\
&=D_0(H_1(t)\otimes H_{ancilla}(t), H_2(t)\otimes H_{ancilla}(t) )&
\nonumber\\
&= D_0(H_1(t), H_2(t))\ .\end{eqnarray}
Thus the time required to distinguish between the two Hamiltonians
eq. (\ref{general})
does not depend on the presence of the ancilla.

The basic reason why an ancilla does not help in distinguishing
between two unknown Hamiltonians is that all the argument of section
\ref{TwoH} depends on the eigenvalues of the difference between the two
Hamiltonians $H^d$ but not on the degeneracy of the
eigenvalues. Including and ancilla does not change the eigenvalues of
$H^d$, but changes their degeneracy.

\subsection{Attaining the bound in the dichotomic case}\label{Att}

We shall now show that one can always attain the bound
eq. (\ref{TDH}). In general this will require 
 putting the particle in a superposition of ``in the
box'' and ``out of the box'' states.
Equation (\ref{psipsi}) shows that to attain the bound the particle must
be
kept in a superposition with equal weights of the eigenstates of
$\tilde H^d= \tilde H_1^d - \tilde H_2^d$ with maximal and minimal eigenvalue,
$|\tilde E^{d}_{max}\rangle$ and
$|\tilde E^d_{min}\rangle$.
In the preceeding section we gave two examples of
how to do this
when the two Hamiltonians $H_1$ and
$H_2$ commute.
When the Hamiltonians do not commute one must use a more complicated
strategy.
Let us first rewrite the Hamiltonians $\tilde H_1$ and $\tilde H_2$ as
\begin{eqnarray}
\tilde H_1 = \tilde H^{+} + \tilde H^d /2 \quad &,&
\quad \tilde H_2 = \tilde H^{+} - \tilde H^d /2 \ ,
\nonumber\\
\tilde H^{+} = ( \tilde H_1 + \tilde H_2 ) /2
\quad &,& \quad  \tilde H^d = \tilde H_1 - \tilde H_2 \ .
\end{eqnarray}
The evolution during a small time interval $\tau$ can then be written as
\begin{eqnarray}
\exp({-i \tilde H_1 \tau})
&=&\exp(-i (\tilde H^{+} + \tilde H^d /2) \tau)\nonumber\\
&\simeq& \exp(-i \tilde H^{+} \tau) \exp(-i \tilde H^d \tau /2)
\exp(O(\tau^2))
 \ ,\nonumber\\
\exp ({-i \tilde H_2 \tau})
&=&\exp(-i (\tilde H^{+} - \tilde H^d /2) \tau)\nonumber\\
&\simeq& \exp(-i \tilde H^{+} \tau) \exp(+i \tilde H^d \tau /2)
\exp(O(\tau^2))  \ .\nonumber\\
\label{15}
\end{eqnarray}
To distinguish the two Hamiltonians we initially
prepare the
system in the state
 $|\psi (0)\rangle = (|\tilde E^d_{max}\rangle + |\tilde E^d_{min}\rangle
 )/\sqrt{2}$.  The evolution is the following.
We let the unknown Hamiltonian
act for a small time $\tau = T/N$ where $T =
\pi / D_0( H_1,H_2)$ and $N$ is a large integer.
We then act on the system with the unitary
transformation $U = \exp(-i\nu \tilde H^d) \exp(+i \tilde H^{+} \tau)$
where $\nu$ is
an
arbitrary real number.
The term on the right in $U$ cancels the term on the left in eq.
(\ref{15}).
This unitary evolution $U$  can in principle be done in
an arbitrarily short time. After $N$ repetitions, the
evolution is
\begin{eqnarray}
|\psi(T) \rangle &=& \left [
\exp(-i(\nu \pm 1/2) \tilde H^d \tau)\exp(O(\tau^2)) \right]^N
|\psi (0)\rangle
\nonumber\\
&\simeq&
\exp (-i \pi(\nu \pm 1/2) \tilde H^d ) |\psi (0)\rangle
\end{eqnarray}
where we have written an expression valid in the limit of large $N$
Thus we obtain two orthogonal states which can be distinguished with
certainty.

Note that if we take the arbitrary real number $\nu$ to be either $\pm
1/2$, then $U = \exp(-i\tau \tilde H_{1,2})$, that is we have exactly
canceled the evolution of one of the Hamiltonians. This is the
technique that is proposed in \cite{CPR}.

\section{Estimating an unknown Hamiltonian}\label{estH}

\subsection{Formulation of the problem}\label{formulation}

In the previous section we considered the situation where one must
{\em distinguish with certainty} between {\em two} possible
Hamiltonians. In the present section we shall consider the problem
where one must {\em estimate with finite precision} what is the
Hamiltonian.

In order to give a precise formulation to this problem, let us suppose that
the possible Hamiltonians are denoted $H_i$ and can occur with a
priori probabilities $p(H_i)$. After an evolution that lasts for a time
$\Delta t$, a measurement is carried out which yields result $j$. The
probability of result $j$ given that the Hamiltonian is $H_i$ is
denoted $p(j|H_i)$. The result of the measurement allows one to
estimate what is the Hamiltonian. We shall suppose that one makes a
guess of what is the true Hamiltonian. Call the guess
$H_j^{guess}$. Note that $H_j^{guess}$ can be one of the initial
Hamiltonians $H_i$, or it could be a different Hamiltonian
altogether. The quality of the guess is measured by an uncertainty
\begin{equation}
\Delta H = \sum_i p(H_i) \sum_j p(j|H_i) D(H_i, H_j^{guess})  \ .
\label{DeltaH}
\end{equation}
In this equation $D(H_i, H_j^{guess})$ is a distance on the space of
Hamiltonians that measures how close the guess is to the true
Hamiltonian $H_i$.  It is natural to normalize the distance
$D$, for instance by requiring that $D(H, H + E \openone) = E$.

There are many different distances on the
space of Hamiltonians.
For instance $D(H_1,H_2) = \sqrt{ Tr (H_1 - H_2)^2}/ d$
where $d$ is the dimension of the Hilbert space. In general the
distance that one will use will depend on the specific problem one
considers. In
the present paper we shall use
the distance $D_0$ introduced in eq. (\ref{DH}). The reason is that
this is the distance that
appears in the results of section \ref{EstH}, and these
results are used in the arguments below.
We expect that using another distance would change quantitatively, but
not qualitatively, our results.

One of the most fundamental questions concerning the estimation of an
unknown Hamiltonian is the relation between the precision $\Delta H$
with which the Hamiltonian is known and the time $\Delta t$ used to
carry out the estimation. The remainder of this section is devoted to
addressing this question.

\subsection{
Estimating a Hamiltonian which can only take two values,
  $H_1$ or $H_2$}\label{estiH1H2}

As a first application of the general problem of estimating an unknown
Hamiltonian consider the particular situation in which their are only
two possible Hamiltonians, $H_1$ and $H_2$, which are equally probably
$(p(H_1)=p(H_2)=1/2$) and the task is to
estimate which it is in a finite time $\Delta t$. The quality of this
estimate shall be expressed by using  as
distance on the space of Hamiltonians the distance $D_0$ defined in
eq. (\ref{DH}).
The
techniques developed in section \ref{EstH} will allow us to solve
this problem exactly.

In section \ref{EstH}  it was shown that if
$\Delta t \geq \pi / D_0(H_1,H_2)$, then the two
Hamiltonians can be distinguished perfectly and therefore $\Delta H$
is zero. On the other hand when $\Delta t\to 0$, it is
impossible to obtain any information about the Hamiltonians and the best
strategy is to randomly guess either $H_1$ or $H_2$. Hence in this
limit $\Delta H \to D_0(H_1,H_2)/2$. For intermediate times $\Delta
H$
will decrease from $D_0(H_1,H_2)/2$ to zero as a function of
$\Delta t$. We shall show that for
the optimal estimation strategy, $\Delta
H$ is given by
\begin{equation}
\Delta H = \max \left\{ 0,
{ D_0(H_1,H_2) \over 2} \left (
1 - \sin ( {D_0(H_1,H_2) \Delta t \over 2}) \right ) \right\} \ .
\label{estimateDicho}
\end{equation}

To prove this let us first show that for an optimal guessing strategy
it is sufficient that the guessed Hamiltonian be either $H_1$ or
$H_2$. We recall that in the estimation problem as formulated in
section \ref{formulation}
we left open the possibility of guessing a Hamiltonian
that is not one of the possible Hamiltonians $H_i$. In the present
case it is not necessary to consider such possibilities.

To show this let us consider the contribution of guess $j$ to $\Delta
H$ (we denote this contribution $\Delta H(j)$):
\begin{eqnarray}
\Delta H (j) &= &{1 \over 2} p(j|H_1) D(H_1,H_j^{guess}) \nonumber\\
& & + {1 \over 2} p(j|H_2) D(H_2,H_j^{guess}) \ .
\label{G1}
\end{eqnarray}
Using the triangle inequality we can write
\begin{eqnarray}
\Delta H (j) &\geq &{1 \over 2} p(j|H_1) D(H_1,H_2) \nonumber\\
& & + {1 \over 2} [p(j|H_2) - p(j|H_1)]D(H_2,H_j^{guess}) \ .
\end{eqnarray}
If $p(j|H_2) - p(j|H_1) \geq 0$ we finally have
\begin{eqnarray}
\Delta H (j) &\geq &{1 \over 2} p(j|H_1) D(H_1,H_2)
\label{G2}
\end{eqnarray}
with equality if and only if $H_j^{guess} = H_2$. Thus if the a
posteriori probability that the Hamiltonian was $H_2$ is greater than
the a posteriori probability that the Hamiltonian was $H_1$ one should
guess that the Hamiltonian is $H_2$. And conversely  if the a
posteriori probability that the Hamiltonian was $H_1$ is greater than
the a posteriori probability that the Hamiltonian was $H_2$ one should
guess that the Hamiltonian is $H_1$.

Let us now consider the optimal evolution and measurement
strategy. The estimation strategy starts with a given quantum
state $\psi_0$. If the Hamiltonian is $H_1$ this state evolves into
$\psi_1 (t)$ whereas if the Hamiltonian is $H_2$ the state evolves
into $\psi_2(t)$. In section \ref{TwoH} it was shown that the overlap
between these two states must obey the inequality
\begin{equation}
|\langle \psi_1(t)|\psi_2(t)\rangle| \geq \cos{D_0 t \over 2}
\end{equation}
with equality attained for the optimal strategy
(we have denoted $D_0 = D_0 (H_1,H_2)$).
Furthermore it is shown in \cite{Helstrom} that the probability $p_E$ of
making an error when trying to distinguish two equiprobable
states $\psi_1$ and
$\psi_2$ is bounded by
\begin{eqnarray}
p_E \geq {1 - \sqrt{1 - |\langle \psi_1|\psi_2\rangle|^2} \over
  2}
\geq {1 - \sin (D_0 t / 2) \over
  2}
\end{eqnarray}
with equality attained if one carries out a von Neumann measurement
of the basis $(|\psi^\parallel\rangle \pm |\psi^\perp\rangle
)/2$. Hence we find that for the optimal strategy
\begin{eqnarray}
\Delta H = p_E D_0
= { D_0 \over 2} \left (
1 - \sin ( {D_0  \Delta t \over 2}) \right )
\end{eqnarray}
which proves eq. (\ref{estimateDicho}).

\subsection{Estimating a completely unknown Hamiltonian}\label{ManyH}

We now consider the situation where the Hamiltonian is completely
unknown. Once more we shall use as distance on the space of
Hamiltonians the distance $D_0$ of eq. (\ref{DH}).
We shall show that in this case the precision $\Delta H$
with which the
Hamiltonian is estimated and the time used to estimate the Hamiltonian
must obey the constraint
\begin{equation}
\Delta H \Delta t \geq {1 \over 4} \ .
\label{uncertainty}
\end{equation}
This constitutes one of the fundamental results of this paper.

To prove this,
we will
contrast two situations. In the first situation (which is the one we
are interested in) the experimenter has
no information about the Hamiltonian.
In the second situation we imagine that there is
a ``spy'' that knows the true Hamiltonian, call it $H_0$. The spy
then tells  the experimenter that the true Hamiltonian is either
$H_0$ or some other Hamiltonian $H_1$. The a priori probabilities that
the spy chooses Hamiltonian $H_0$ or $H_1$ are equal.

Denote by $\Delta H$ the maximum precision with which the Hamiltonian can be
known in the first case and by
$\Delta H_{dicho}$ the maximum precision with which the Hamiltonian can be
known  with the help of the spy. Obviously $ \Delta t \Delta H \geq
\Delta t\Delta
H_{dicho}$ since the time intervals are the same in the two situations
and the information provided by the spy can only increase
the precision with which one can estimate the Hamiltonian. This means
that
$\Delta t \Delta H \geq \max
\Delta t\Delta
H_{dicho}$ where the maximum is taken over all possible choices of the
spy.
The results of section \ref{estiH1H2} can be used to show that
$ \max
\Delta t\Delta
H_{dicho} \geq 1/4$, which proves  eq. (\ref{uncertainty}).

To show that $ \max
\Delta t\Delta
H_{dicho} \geq 1/4$, note that in the dichotomic case, the
product $\Delta H \Delta t$ takes the form
\begin{equation}
\Delta H \Delta t = {D_0 \Delta t \over 2}
\left( 1 -  \sin ( {D_0  \Delta t \over 2})\right)
\label{prod}
\end{equation}
 for $0 \leq \Delta t \leq \pi / D_0$. For small times  this tends to
 zero since $\Delta H$ is bounded and $\Delta t \to 0$. And for
 $\Delta t \geq \pi / D_0$ the product is zero since $\Delta
 H=0$. There is an intermediate time where the product attains its
 maximum. One easily shows (using $\sin x \leq x$ for $x \geq 0$,
 which implies that $x (1-\sin  x) \geq x (1-x)$) that the maximum
 value of this product is greater than a quarter:
\begin{equation}
\max  \Delta H \Delta t \geq {1 \over 4}\ .
\label{maxDHDT}
\end{equation}

\subsection{Estimating a completely unknown Hamiltonian acting in a
  $d$ dimensional space}

Equation (\ref{uncertainty}) gives a lower bound on the product of
 the precision with which a
completely unkown Hamiltonian is measured and the time taken to
estimate it. We beleive that 
this lower bound is not tight and that in general 
a stronger lower bound should hold. We do not know at present
what form this stronger lower bound will take, but we beleive that
it should depend on the dimensionality of the Hilbert space on which
the Hamiltonian acts. 

An indication that this should be the case is provided 
 by an
example due to Farhi and Gutmann \cite{FG} inspired by Grover's search
algorithm\cite{G}. In this example one must distinguish between $d$
Hamiltonians of the form $H_k = E |k\rangle \langle k |$ where the $d$
states $|k\rangle$ form an orthonormal basis. Farhi and Gutmann
 show that in order to
distinguish these Hamiltonians perfectly, a minimum time of 
$\Delta t \geq c d^{1/2} / E$ is necessary (where $c$ is some positive
constant). 

This example shows that
there are situations where estimating an unkown Hamiltonian becomes
increasingly difficult as the dimension $d$ of the Hilbert space on which
it acts increases. However in the Fahri--Gutmann example, the unkown
Hamiltonian has a very specific form which is known before hand. We
have obtained preliminary indications that when 
the Hamiltonian is completely unkown, estimating it should take
substantially more time than sugested by the Fahri-Gutmann example.
We hope to report on this issue in a future publication.

\section{Measuring Energy When the Hamiltonian is unknown}\label{MeasE}

\subsection{Introduction}\label{FormulE}

The results presented in the preceding section
concerning the estimation of Hamiltonians
have important implications for
energy measurements. As shown in \cite{AB} the
energy of the state can be measured in an arbitrarily short
time. However a
careful scrutiny of the arguments of  \cite{AB} shows that a quick energy
measurement is possible only if the Hamiltonian is known. In the
example discussed in the introduction, it is possible to carry out a quick
energy measurement only because we know that the particle is in a
magnetic field $\vec B_0$ of known magnitude
pointing along the $+z$ direction. Suppose however that the magnetic
field is pointing initially either along the $+z$ or the $-z$
axis. Then a measurement of $\sigma_z$ yields no information about the
energy.
Thus in order to
determine the energy of the particle, we  must also
determine the magnetic field. That is we must also determine
what is the Hamiltonian. But as we discussed above, determining the
Hamiltonian will take a minimum time $\Delta t$. (We suppose that the
only way we can
probe the
magnetic field is with a particle of magnetic moment $\mu$. Of course
if we could use a particle of larger magnetic moment, the measurement
of the magnetic field could be done faster). Thus in this example
measuring the energy of the state cannot be done instantaneously
because the Hamiltonian is not perfectly known.

In fact this is a very general result. We shall show below that {\em if the
Hamiltonian acting on a system is completely unknown,
then the precision $\Delta E$ with which one can estimate
the energy of the state in a time interval
$\Delta
t$ obeys the constraint}
\begin{equation}
\Delta t \ \Delta E \geq 1/4 \ .
\label{DtDE}
\end{equation}

This assertion follows easily from the results obtained  in the
previous section. However before proving it
we first need to define with precision what
we mean by accuracy $\Delta E$ of an energy measurement.

\subsection{Accuracy of an energy measurement}\label{AccEn}

An ideal energy measurement is a von Neumann measurement of the
Hamiltonian operator $H = \sum_E E |E\rangle\langle E|$. If the
quantum state is $|\psi\rangle$, the measurement gives result $E$ with
probability $p(E|\psi) = |\langle \psi |E\rangle|^2$.

Let us consider an imperfect measurement. This measurement will
predict that the energy is $E'$ with probability $p(E')$. Neither
the energies $E'$ nor the probabilities $p(E')$ need coincide with
the energies and probabilities for an ideal energy
measurement. Nevertheless we would like to define in a precise way the
accuracy of an energy measurement.

The simplest situation in which to define  the accuracy of an energy
measurement is when the quantum state is an energy eigenstate
$|\psi\rangle = |E\rangle$. In this case the true energy of the state
is well defined. Hence the accuracy of the imperfect energy
measurement is simply the amount by which the energies $E'$ differ
from the true energy $E$:
\begin{equation}
\Delta E = \sum_{E'} p(E'|E) |E'-E| \label{defDE}
\end{equation}
where $p(E'|E)$ is the probability that the estimated energy
is $E'$ when the quantum state is $\psi = |E\rangle$.

 If the state is not an energy eigenstate then we define the
accuracy of the energy measurement as the average over the 
probability $|\langle E|\psi\rangle|^2$ that an ideal energy
measurement gives result $E$ times of the accuracy of the
measurement if the state is $|E\rangle$:
\begin{equation}
\Delta E = \sum_E  |\langle E|\psi\rangle|^2 \sum_{E'} p(E' |E)
|E'-E|\ . \label{defDE2}
\end{equation}

\subsection{Proof of the time-energy uncertainty relation for energy
  measurements when the Hamiltonian is completely unknown}\label{proofDE}

We start by noting that the proof of eq. (\ref{DtDE}) when $\psi$ is
not an eigenstate of the Hamiltonian follows from the case where
$\psi$ is an eigenstate of the Hamiltonian since we have defined in
(\ref{defDE2}) the uncertainty when $\psi$ is
not an eigenstate of the Hamiltonian as the average of the
uncertainties when $\psi$ is
an eigenstate of the Hamiltonian times the probability that a
measurement of the Hamiltonian operator yields the corresponding
energy.
Therefore we can restrict ourselves to considering the case where
$\psi =|E\rangle$ is an eigenstate of the Hamiltonian $H$. Since the
Hamiltonian $H$ is completely unknown, the state $\psi$ is also
unknown.

To prove eq. (\ref{DtDE}) we fix $\Delta t$ and contrast
as in the section \ref{ManyH}, two
situations. In the first one has no information about the
Hamiltonian. In the second a spy gives the additional information that
the Hamiltonian is either $H_0$ (the true Hamiltonian) or $H_1$. We
shall suppose that $H_1 = H_0 + \epsilon \openone$ where $\epsilon$ is a
c-number and $\openone$ is the identity operator. We shall further
suppose that with the information provided by the spy there is equal a
priori probabilities that the Hamiltonian is $H_0$ or $H_1$. Let us
denote the energy uncertainty in the first situation by $\Delta E$ and
in the second situation by $\Delta
E_{spy}$. Obviously we have $\Delta t \Delta E \geq \Delta t \Delta
E_{spy}$ since $\Delta t$ is the same in both situations and the
information provided by the spy can only decrease the energy
uncertainty. We therefore want to put a bound on $\Delta
E_{spy}$.

First note that since the two Hamiltonians $H_0$ and $H_1 = H_0 +
\epsilon \openone$ commute the experimenter can immediately determine
what is the state $\psi$ by measuring the operator $H_0$.  This
measurement can in principle be done arbitrarily fast.
Hence the experimenter knows that the energy is either $E$, the true
energy, or $E+\epsilon$.

Since the experimenter has only two possibilities between which to
choose, an optimal strategy will consist of guessing either that
the energy is $E$ or $E+\epsilon$. It is not necessary to consider
other possibilities such as guessing that the energy is $E +
\epsilon /2$. Furthermore the energy uncertainty will be $\Delta
E_{spy} = \epsilon p_E$ where $p_E$ is the probability of making
the wrong guess. The proof of these assertions follows from the
fact that $\Delta E$, as defined in eq. (\ref{defDE}), is linear
in the probabilities $p(E')$ times a distance $|E'-E|$ on the
space of energies. Hence the arguments of section \ref{estiH1H2},
eqs \ref{G1} to \ref{G2}, can be used in the present case.

But the error probability $p_E$ of mistaking one energy for the other
is identical to the error probability of wrongly identifying the
Hamiltonians $H_0$ and $H_1$ (since knowing the energy is equivalent
to knowing the Hamiltonian). Hence in the present case $\Delta E_{spy}
= \Delta H$ where $\Delta H$ is the uncertainty in estimating the two
Hamiltonians $H_0$ and $H_1$. But we have shown in section
\ref{estiH1H2} that for given $\Delta t$ there exists a choice of
$\epsilon$ such that $\Delta t \Delta H \geq 1/4$.

\subsection{Saturating the time-energy uncertainty for energy
  measurements when the Hamiltonian is unknown?}

 We now address the question of {\em whether it is
possible
to devise a universal measurement
strategy that can determine the energy of an
unknown  state even if there is no prior knowledge about the Hamiltonian?}
Such universal measurement strategies exist and are well known.
We illustrate them by
a typical example, namely the emission of electromagnetic
radiation by an excited state of an atom.

We then inquire whether this
measurement strategy saturates the time-energy uncertainty relation
obtained in the previous section. It turns out that for this
measurement strategy, and  using the
definition
of section \ref{AccEn}, $\Delta t \Delta E $ is infinite. However in a
qualitative way this measurement strategy does obey a time energy
uncertainty. This is discussed in detail.

As mentioned in the introduction the
emission of one or more photons by an excited atomic state can be
viewed as a measurement of the energy of the electrons. The coupling
between the measuring apparatus (the electromagnetic field $A^\mu$ ) and the
system (the electron) is realized through the interaction
\begin{equation}
H_{int}=\int d^3x  A^\mu(x) J_\mu(x)
\label{Hint}
\end{equation}
 where $J_\mu(x)$ is the electric current.
In interaction representation it takes the form
\begin{equation}
H_{int} = \int_0^\infty d\omega
\sum_k (a^\dagger_{\omega k} e^{i\omega t} +
a_{\omega k} e^{-i\omega t} ) J_{k\omega} \ .
\label{elect}
\end{equation}
Here $\omega$ is the energy of the photons, $k$ represents
other degrees of freedom of the photons in addition to their energy
(momentum and polarization) and $J_{k\omega}$ are operators
acting on the electrons Hilbert space. The photons are taken to be
initially in their ground state $a_{\omega k} |0\rangle =0$.

The interaction Hamiltonian  eq. (\ref{Hint}) is
independent of the electron Hamiltonian, ie. it
is independent of whether the electron is bound to a proton, a Helium
nucleus, a molecule, etc\ldots. Therefore such a measurement
can determine the
energy of an unknown state, independently of the Hamiltonian. It can
also determine what is the Hamiltonian, since the energy of the
emitted photons will
differ if the electron is bound to a proton, a Helium nucleus, a
molecule,
etc\ldots. The
price to pay for this universality is that the energy resolution of
the measurement and the time necessary for the measurement
are constrained by the time-energy uncertainty.
The arguments presented in this paper show that this will always be
the case for a measurement of energy which does not take into account
prior knowledge about the Hamiltonian.

 An important limitation of the above measurement scheme is
that the emitted photon only reveals the difference in energy
between the initial and final state of the atom. If there are
several allowed transitions with identical energy differences,
then the measurement will not allow these initial states to be
differentiated. It would have to be complemented by a second
measurement to determine which of the possible final states the
atom reached. Nevertheless the important point of this example is
to show that in principle it is possible to come close to
saturating the time energy uncertainty relation when estimating
the energy of a system whose Hamiltonian is unknown.

In order to see how close we come to saturating the time-energy
uncertainty in this scheme, let us examine it in more detail. The
probability density that a photon is emitted at time $t$ is:
\begin{equation}
P(\mbox{decay at time $t$})= \gamma e^{- \gamma t} \ .
\end{equation}
Thus the time it takes to complete the measurement is not well
defined. Rather this time is variable but its mean is finite
\begin{equation}
\int dt \ t  \  P(\mbox{decay at time $t$})= \gamma^{-1}\ .
\end{equation}
This is to be contrasted with the situation envisaged in the previous
sections where we required that the measurement be finished after some
time interval $\Delta t$.

Let us now consider the energy of the emitted photon. If the true
energy of the electronic state is $E_0$, then the probability
density that the emitted photon has energy $E$ is
\begin{equation}
P(\mbox{emitted photon has energy $E$})=
{1 \over \pi} {\gamma \over  \gamma^2 + (E- E_0)^2}\ .
\end{equation}
If we compute the accuracy of the energy measurement using the
definition eq. (\ref{defDE2}), we find
\begin{eqnarray}
\Delta E &=& \int dE  P(\mbox{emitted photon has energy $E$})|E -E_0|
\nonumber\\
&=& +\infty \ .
\end{eqnarray}

Thus this measurement satisfies the time energy uncertainty
relation. In fact the product $\Delta t \Delta E$ is infinite since
both the time it takes to complete the measurement and the energy
uncertainty are infinite.
However we note that if we
modify the definition of $\Delta t$ to be the mean time
$\gamma^{-1}$
taken to carry out the measurement, and if we modify the definition of
the energy uncertainty $\Delta E$ to be the
linewidth  $\gamma$, then this measurement does obey a time
energy uncertainty relation.

Therefore it may be possible to devise a
better energy measurement that saturates the time-energy uncertainty
derived in section \ref{proofDE}. Or this uncertainty relation is too strong,
and one can prove a weaker form of the uncertainty relation, for
instance using as definition of  $\Delta t$ and  $\Delta E$ the mean
time taken to do the measurement and the line width. In the latter
case the measurement just described would be optimal in the sense that
it would saturate the time energy uncertainty for energy measurements
when the Hamiltonian is unknown.

\subsection{Estimating energy when one has partial knowledge about the
  Hamiltonian}\label{partialK}

In the previous section we considered the situation where one wants to
estimate the energy of an unknown state but one has no
prior knowledge about the Hamiltonian. When some prior knowledge is
available the situation is considerably more complicated and the
relation between the time used for the measurement and the precision
with which the energy can be estimated will depend on the details of
the problem.

To illustrate
this we consider two examples. First consider the case of two
Hamiltonians $H_1$ and $H_2$ that have the same eigenstates $H_1
\psi_k = E_{1k} \psi_k$,  $H_2
\psi_k = E_{2k} \psi_k$ and their eigenvalues coincide except for one
eigenstate: $E_{1k}=E_{2k} (k\neq k_0)$ but $E_{1k_0}\neq
E_{2k_0}$. Suppose we must determine the energy of an unknown state
$\Psi$. A strategy to do this in a short time is to first carry out a
von Neumann measurement of the basis $\psi_k$ that diagonalizes $H_1$
and $H_2$ (this can be done in an arbitrarily short time). If one finds
that the outcome $k$ is different from $k_0$, then one immediately
knows the energy. On the other hand if $k=k_0$, then to know the
energy one must determine what is the Hamiltonian. This takes a time
$\Delta t = \pi / | E_{1k_0}- E_{2k_0}|$. If the unknown state $\Psi$
was uniformly distributed in Hilbert space (denoted ${\cal H}$),
then the probability that
$\Psi$ belongs to subspace $k_0$ is $1/ \mbox{dim}{\cal H}$ and the
average time necessary to determine the energy of the state is
$\pi /   \mbox{dim}{\cal H} | E_{1k_0}- E_{2k_0}|$ which is much
smaller than the time needed to determine the Hamiltonian. The reason
for this difference in time scales is because the particle has most of
its support in a part of the Hilbert space where the two Hamiltonians
do not differ.

Our second example is superficially similar to the previous one. But
more careful consideration show some subtle differences. We consider
a particle confined to a (one dimensional) box. The potential in the box
vanishes everywhere, except in a corner where it may be either zero or
take a large negative value. In order to measure the energy of the
particle in a minimum time
the following strategy seems natural. First we measure whether the
particle is in the corner or not. (This position measurement should be
slightly fuzzy so as not to disturb the momentum  too much).
This measurement can in principle be
done in arbitrarily short time. If the particle is not in the corner,
we measure its momentum, and hence know its energy.
This can also be done in arbitrarily short
time. If the particle is in the corner, then we must determine the
value of the potential in order to know the energy of the
particle. This takes a finite time of order $1/ \Delta V$, where
$\Delta V$ is the uncertainty in the potential.

Thus in this case it
seems that the
minimum time required for the energy measurement depends essentially
on the probability of the particle being in the corner of the box
rather
than on
the precision $\Delta E$ with which one wants to know the energy.
However the situation is more complicated. The above procedure
approximates to some extent a von Neumann measurement of the
Hamiltonian operator. Indeed the statistics of the measurement
outcomes are such that they reproduce correctly the moments of the
Hamiltonian operator. Thus for instance upon repeating the measurement
many times one will obtain a good estimate of
the average energy
$\langle H \rangle$, or the average value of any power of the
Hamiltonian $\langle H^n \rangle$.
However the above procedure is not equivalent to a von Neumann
measurement of the Hamiltonian operator (which is the task we set out
to perform). Indeed there
are
some functions of the Hamiltonian which cannot be estimated correctly
with the above procedure.

Suppose for instance that the particle is known to be approximately
localized at
a distance $d$ from the corner of the box ($d$ is taken to be much
larger than the size of the region where the potential is unknown) and
suppose that it is known that the momentum of the particle is
approximately $p$. Since the particle is far from the corner,
by measuring the momentum of the particle one has some information
about its energy. For instance independent repetitions of the
measurement will yield  estimates of the moments of the Hamiltonian
operator
$\langle H^n \rangle$. Such measurements of the momentum can be done
in an arbitrarily short
time. However suppose that one wants to measure the operator
$\cos (m d H / p) = ( e^{im d H / p} + e^{-im d H / p} )/2$ (where $m$
is the mass of the particle). This is the real part of the operator
that evolves the particle from its initial position up to the corner
where the potential is unknown. The expectation value of the above
operator
clearly depends on the value of the potential in the corner, and
therefore it can only be determined in a time of order $1/V$.

The origin of this surprising situation is that in the above example the exact
spectrum of $H$ depends on the potential in the corner. When the
particle is far from the corner, most questions concerning the energy
of the particle are independent of the exact spectrum of $H$. But a
variable such as $\cos (m d H / p)$ is sensitive to the exact
spectrum and therefore in order to measure it one must know what is
the potential in the corner.
Operators such as $\cos (m d H / p)$ are called modular
variables, and have been introduced in \cite{modular}.

\section{Conclusions}

In the present article we have shown that if the Hamiltonian that
governs the evolution of a quantum
system is unknown, then the time necessary to estimate the Hamiltonian
obeys a time-energy uncertainty relation eq. (\ref{TDH}). To this end
 we first gave a simple proof of the problem considered by Childs,
Preskill and Renes\cite{CPR} where there are only two possible
Hamiltonians between which one must choose. We then showed how to
extend this result to the case where there are many Hamiltonians among
which one must choose. The bound we obtained is probably not tight
when the unknown Hamiltonians act in a space of large dimensionality
$d>2$, 
and it should be possible to refine it by a more detailed
analysis.

Our results concerning the time-energy uncertainty relation applied to
estimating Hamiltonians have many applications. In particular they
provide new insight about how the time-energy uncertainty applies to
energy measurements. It has been shown by Aharonov and Bohm
that if the Hamiltonian of
the system is known, then the Hamiltonian can in principle be measured
in
arbitrarily short time. On the other hand we show that
 if the Hamiltonian is unknown then the
energy measurement cannot be done in arbitrarily short time. The
minimum amount of time required depends on the details of the problem,
for instance what is the prior knowledge about the Hamiltonian, what
is the prior knowledge about the state, and exactly what one wants to
know about the energy of the state. We show that if one has no prior
information about the Hamiltonian, then the time taken to carry out
the measurement and the precision with which the energy is measured
obey a time-energy uncertainty relation.

We also show that one can
devise a measurement of the energy of a quantum system that always
works, independently of any prior knowledge about the system. Such a
measurement is obtained by coupling the system to an
external apparatus that oscillates at all frequencies and such that
each frequency is coupled to different
degrees of freedom of the apparatus. This is illustrated in
eq. (\ref{elect}) in the case where the external apparatus is taken to
be the electromagnetic field. Such measurements do not saturate our
time energy uncertainty relation, although they do obey a
qualitatively similar uncertainty relation between the life time of
the state and the line width.

To conclude, we find that the real meaning of energy in quantum
mechanics is that of governing the time evolution of a system. To
measure the energy one has to determine the time evolution, and
this takes time. Thus energy measurements require time, and their
precision is limited by the time we have at our disposal. 
On the other hand, in the examples presented by
Aharonov and Bohm \cite{AB}, the Hamiltonian is known in advance,
hence one need not spend time to determine the time-evolution.
Instead, one could find out the value of the energy not by
determining the time evolution, i.e., not by measuring the actual
energy, but by measuring an operator (the operator to which the
Hamiltonian is equal) whose numerical eigenvalue is equal to that
of the energy. However, we emphasize that although this procedure
does yield the numerical values equal to that of the energy, it is
not a proper energy measurement. Indeed, if we believe the
Hamiltonian to be $H$ but in reality it is different, say $H'$,
then the value obtained by the instantaneous Aharonov-Bohm
measuring procedure (which tells us to measure the operator $H$)
would no longer be correct, and, furthermore, we would not know
that our measurement is wrong. Thus a proper energy measurement
necessarily probes the time evolution and therefore cannot be done
instantaneously. Rather the time taken to carry out the
measurement and the precision with which one knows the energy are
constrained by a time energy uncertainty relation $\Delta t \Delta
E \geq 1$.

 \begin{acknowledgments}
  Y.A. acknowledges the support of Grant
471/98 of the Israel Science Foundation established by the Israel
Academy of Sciences and Humanities. S.M. is a ``research
associate'' of the Belgium National Research Fund. S.M. and S.P
acknowledge funding by  project EQUIP (IST-FET program) of the European
Union.
\end{acknowledgments}

\appendix
\section{}

In this appendix we show that
$D_0(H_1,H_2)$ defined in eq. (\ref{DH}) is a distance on the space of
Hamiltonians.
That is $D_0$ is
1) positive:
$D_0 (H_1,H_2) \geq 0$ with equality if and only if $H_1=H_2$; 2)
$D_0$ is symmetric: $D_0 (H_1,H_2)= D_0 (H_2,H_1)$; and 3)
$D_0$ obeys the triangle inequality: $D_0(H_1,H_2) \leq D_0(H_1,H_3) +
D_0(H_3,H_2)$.

Let us first introduce a norm
\begin{equation}
||H||_0 = \max \{ E^{max} - E^{min} , |E^{max}|, |E^{min}|\}
\end{equation}
where $E^{max}, E^{min}$ are the largest, smallest eigenvalues of $H$.
Now
\begin{equation}
D_0 (H_1,H_2) = ||H_1 -H_2||_0\ ,
\end{equation}
hence if we can prove that $||H||_0$ is indeed a norm, then it follows
immediately that $D_0$ is a distance.

We recall that a norm obeys must satisfy the following properties:
1) positivity: $||H||\geq 0$ with equality if and only if $H=0$;
2) linearity: $|| \lambda H || = |\lambda| ||H||$ for any c-number
$\lambda$;
3) triangle inequality $||H_1 + H_2|| \leq ||H_1|| + ||H_2||$.

Properties 1) and 2) are immediate. Let us consider property 3.
Let us denote by $E_1^{max}$ and  $|\psi_1^{max}\rangle$ the largest
eigenvalue of $H_1$ and the corresponding eigenvector;
by $E_2^{max}$ and  $|\psi_2^{max}\rangle$ the largest
eigenvalue of $H_2$ and the corresponding eigenvector;
by $E_{12}^{max}$ and  $|\psi_{12}^{max}\rangle$ the largest
eigenvalue of $H_1+ H_2$ and the corresponding eigenvector.
Let us show that $E_{12}^{max} \leq E_1^{max} + E_2^{max}$.
We have
\begin{eqnarray}
E_{12}^{max} &=& \langle\psi_{12}^{max}| (H_1 + H_2)|\psi_{12}^{max}\rangle
\nonumber\\
&=&\langle\psi_{12}^{max}| H_1 |\psi_{12}^{max}\rangle
+
\langle\psi_{12}^{max}|  H_2|\psi_{12}^{max}\rangle
\nonumber\\
&\leq&
\langle\psi_{1}^{max}| H_1 |\psi_{1}^{max}\rangle
+
\langle\psi_{2}^{max}|  H_2|\psi_{2}^{max}\rangle \nonumber\\
&=&  E_1^{max} + E_2^{max} \ .
\end{eqnarray}
Similarly we have
$E_{12}^{min} \geq E_1^{min} + E_2^{min}$ where $E_{12}^{min},
E_1^{min},
E_2^{min}$ are the smallest eigenvalues of $H_1 +H_2, H_1, H_2$
respectively.
The triangle inequality follows from these relations between
eigenvalues and from the definition of $||H||_0$.


\begin{references}

\bibitem{LP} L. D. Landau and R. Peierls, Z. Physik {\bf 69} (1931) 56

\bibitem{LL} L. Landau and E. Lifschitz, {\em Quantum Mechanics}
  (Pergamon Press, New York, 1958), pp. 150-153

\bibitem{AB} Y. Aharonov and D. Bohm, Phys. Rev. {\bf 122} (1961) 1649

\bibitem{AR} Y. Aharonov and B. Reznik, Phys. Rev. Lett. {\bf 84} (2000)
1368

\bibitem{CPR} A. M. Childs, J. Preskill, J. Renes, J. Mod. Optics,
  {\bf 47} (2000) 155

\bibitem{MM} G. Mitchison and S. Massar,  Phys, Rev. A {\bf 63} (2001),
032105

\bibitem{BBBV} C. H. Bennett, E. Bernstein, G. Brassard, and
  U. V. Vazirani, SIAM Journal on Computing 26, 1510 (1997),
  quant-ph/9701001

\bibitem{FG} E. Farhi and S. Gutmann, Phys. Rev. A {\bf 57}, 2403
  (1998), quant-ph/9612026

\bibitem{G} L. K. Grover, {\em A fast quantum mechanical algorithm
for database search}, in Proc. 28th Annual ACM Symposium on the
Theory of Computing, 212 (ACM Press, New York, 1996),
quant-ph/9605043

\bibitem{Helstrom} C. W. Helstrom, {\em Quantum Detection and Estimation
Theory}, (Academic Press, New York,1976)


\bibitem{BM} D. Bruss, C. Macchiavello,  Phys.Lett. A253 (1999) 249-251

\bibitem{modular} Y. Aharonov, H.
Pendleton and A. Petersen, Int. Jour. of Theor.
Physics, Vol.2, No.3, 213 (1969)


 \end{references}
\end{document}